\documentclass{emulateapj}
\newcommand{\vcs}[1]{\mbox{\boldmath{$\scriptstyle{#1}$}}}
\newcommand{\vc}[1]{\mbox{\boldmath{$#1$}}}
\newcommand{\de}{\mathrm{d}}
\newcommand{\dpa}{\partial}

\newcommand{\nab}{\vc{\nabla}}

\newcommand{\ii}{\mathrm i}
\newcommand{\dv}{\vc{ \xi}}
\DeclareMathSymbol{\varOmega}{\mathord}{letters}{"0A}
\DeclareMathSymbol{\varSigma}{\mathord}{letters}{"06}
\DeclareMathSymbol{\varPsi}{\mathord}{letters}{"09}
\DeclareMathSymbol{\varPhi}{\mathord}{letters}{"08}

\newcommand{\Eq}[1]{equation (\ref{#1})}
\newcommand{\Eqs}[2]{equations (\ref{#1}) and~(\ref{#2})}
\newcommand{\Eqss}[2]{equations (\ref{#1})--(\ref{#2})}

\newcommand{\Sec}[1]{Sect.~\ref{#1}}

\newcommand{\Fig}[1]{Fig.~\ref{#1}}
\newcommand{\Figs}[2]{Figs.~\ref{#1} and \ref{#2}}
 
\newcommand{\Tab}[1]{Table \ref{#1}}

\newcommand{\ts}{\tau_{\rm f}}
\newcommand{\ep}{\epsilon}
\slugcomment{Accepted for publication in ApJ}

\shorttitle{Linear Evolution of Streaming Instabilities}
\shortauthors{Youdin \& Johansen}

\begin{document}

\title{Protoplanetary Disk Turbulence Driven by the Streaming Instability:\\
Linear Evolution and Numerical Methods}

\author{A. Youdin\altaffilmark{1}}
\affil{Princeton University Observatory, Princeton, NJ 08544}
\email{youd@cita.utoronto.ca}

\and

\author{A. Johansen\altaffilmark{2}}
\affil{Max-Planck-Institut f\"ur Astronomie, 69117 Heidelberg, Germany}
\email{johansen@mpia.de}

\altaffiltext{1}{Current address: Canadian
Institute for Theoretical Astrophysics, University of Toronto, 60
Saint George Street, Toronto, ON M5S 3H8, Canada}
\altaffiltext{2}{Visiting Astronomer, Department of Astrophysics, American
Museum of Natural History}

\begin{abstract}

  We present local simulations that verify the linear streaming instability
  that arises from aerodynamic coupling between solids and gas in
  protoplanetary disks. This robust instability creates enhancements in the
  particle density in order to tap the free energy of the relative drift
  between solids and gas, generated by the radial pressure gradient of the
  disk. We confirm the analytic growth rates found by Youdin \& Goodman (2005)
  using grid hydrodynamics to simulate the gas and, alternatively, particle and
  grid representations of the solids. Since the analytic derivation
  approximates particles as a fluid, this work corroborates the streaming
  instability when solids are treated as particles. The idealized physical
  conditions -- axisymmetry, uniform particle size, and the neglect of vertical
  stratification and collisions -- provide a rigorous, well-defined test of any
  numerical algorithm for coupled particle-gas dynamics in protoplanetary
  disks. We describe a numerical particle-mesh implementation of the drag
  force, which is crucial for resolving the coupled oscillations. Finally we
  comment on the balance of energy and angular momentum in two-component disks
  with frictional coupling. A companion paper details the non-linear evolution
  of the streaming instability into saturated turbulence with dense particle
  clumps. 

\end{abstract}


\section{Introduction}

Solid bodies in protoplanetary disks lose angular momentum as they encounter
the headwind of the pressure-supported gas disk. The subsequent radial drift is
fastest for marginally coupled solids whose aerodynamic stopping times are
comparable to the local orbital time \citep{Weidenschilling1977}. For standard
disk models, cm-sized particles at $30$ AU and m-sized bodies at 1 AU suffer
drift times of only approximately $10$ or $100$ orbital periods, respectively.
Rapid infall imposes severe time-scale constraints on the growth into km-sized
solid bodies, or planetesimals, by coagulation. Concerns about the inefficiency
of sticking for macroscopic solids \citep{Benz2000} has also contributed to the
concept of a ``meter-size barrier'' in planet formation \citep[which should not
be misinterpreted as implying that growth to meter sizes is easy, see
e.g.][]{BlumWurm2000}.

The gravitational instability hypothesis \citep{saf69,gw73} postulates that a
sedimented mid-plane layer of small particles (perhaps mm-sized to match
chondrules) will fragment directly into gravitationally bound planetesimals,
avoiding the problems with sticking efficiency and drift. However, disk
turbulence acts to diffuse particles, inhibiting both their vertical settling
to the midplane \citep{wc93,Dubrulle+etal1995} and their ability to collapse
into bound structures \citep{y05b}. Even in a completely laminar disk, particle
settling generates vertical shear in the orbital motion of the gas. This shear
in turn triggers modified Kelvin-Helmholtz instabilities that develop into
turbulence, restricting further sedimentation \citep{gw73,stu80,cdc93}. This
self-induced turbulence may not be able to prevent gravitational collapse if
the solids-to-gas ratio is enhanced above Solar abundances
\citep{sek98,ys02,gl04,stu06}, possibly due to photoevaporation of the gas-rich
surface layers of the stratified disk \citep{tb05} or to pile-ups of solids in
the inner disk from particles that drift in more rapidly from the outer disk
\citep{yc04}. Significant progress has been made in understanding the
turbulence generated by particle settling \citep{is03,go05,jhk06}. However a
simulation that incorporates the full 3D nature of these non-axisymmetric
instabilities, with radial shear and the independent evolution of solids and
gas, has not yet been performed.

This paper addresses the related streaming instability \citep[][ hereafter
referred to as YG]{yg05} where vertical gravity is ignored in order to focus on
a simpler manifestation of particle-gas coupling in Keplerian disks. With no
vertical shear present, the streaming instability is driven by the relative
motion between solids and gas, which is predominantly radial for tightly
coupled particles. The ultimate energy source, as with vertical shear
instabilities, is the radial gas pressure gradient. Particle feedback on gas
dynamics is important not just for establishing the (unstable) equilibrium, but
also for generating escalating oscillations. Consequently, streaming
instabilities trigger exponential growth of arbitrarily small particle density
perturbations, as shown by YG. The single-fluid treatment of \citet{gp00}
discovered a related boundary layer drag instability in stratified disks that
could also concentrate particles. \citet{jhk06} found significant particle
clumping in studies of Kelvin-Helmholz instabilities with particle feedback on
the gas, which those authors hypothesized was a manifestation of non-linear
streaming instabilities. The current study, including a companion paper
\citep{jy07}, explores the consequences of streaming instabilities, and more
generally the role of particle-gas coupling in protoplanetary disks. This paper
demonstrates that our simulations faithfully reproduce the linear physics of
the streaming instability, whether the solids are modeled as a fluid or
Lagrangian particles.

The paper is built up as follows. In \S\ref{s:eqns} we present the basic
equations of our dynamical system and review the streaming instability. Section
\ref{s:numerics} describes the numerical methods, including the communication
of drag forces between particles and a grid in \S\ref{s:dragcalc}. Our main
results, in \S\ref{s:linear}, numerically confirm the linear streaming
instability. In \S\ref{s:EL} we analyze energy and angular momentum balance in
a coupled two-fluid system. We discuss our results in \S\ref{s:disc}. The
appendices contain an analysis of interpolation and assignment errors in
different particle-mesh approaches to calculating drag forces (Appendix
\ref{s:Werrors}), a non-axisymmetric analytical problem used to test drag force
assignment over shear-periodic boundaries (Appendix \ref{s:shearing}), and a
recipe to minimize Poission noise in seeding linear particle density
perturbations (Appendix \ref{s:coldstart}). A companion paper, Johansen \&
Youdin (2007, hereafter referred to as JY), describes the full non-linear
evolution of the streaming instability into turbulence.

\section{Streaming Instability: Analytics}\label{s:eqns}

\subsection{Basic Equations}

We describe the local dynamics of the gas and solid component of a
protoplanetary disk in the shearing sheet approximation
\citep[e.g.][]{glb65II,GoldreichTremaine1978}. The Cartesian coordinate frame
corotates with the Keplerian frequency $\varOmega$ at an arbitrary orbital
distance $r$ from the central gravity source. The coordinate axes are oriented
such that $x$ points radially outwards, $y$ points along the rotation direction
of the disk, while $z$ points vertically out of the disk, parallel to the
Keplerian rotation vector $\vc{\varOmega}$. Our unstratified model omits
vertical gravity. We measure all velocities relative to the linearized
Keplerian shear flow in the rotating frame $\vc{V}_0 = V_{y,0}\hat{\vc{y}} =
-(3/2) \varOmega x \hat{\vc{y}}$.

\subsubsection{Solids as a Fluid}

Analytic investigations are greatly simplified by treating solid particles as a
continuous fluid of density $\rho_{\rm p}$ and velocity $\vc{w}$, which evolve
according to shearing sheet equations of continuity and motion
\begin{eqnarray}
  \frac{\dpa \rho_{\rm p}}{\dpa t} + \vc{w} \cdot \nab \rho_{\rm p}
     - \frac{3}{2} \varOmega x \frac{\dpa \rho_{\rm p}}{\dpa y}
     &=&  - \rho_{\rm p}\nab \cdot \vc{w} \, ,
  \label{eq:dustcontinuity}\\
  \frac{\dpa \vc{w}}{\dpa t} + (\vc{w}\cdot\nab)\vc{w} 
      - \frac{3}{2} \varOmega x \frac{\dpa \vc{w}}{\dpa y}
     & =& 2 \varOmega w_y \hat{\vc{x}} \nonumber \\
     &  & \hspace{-1.0cm} - \frac{1}{2} \varOmega w_x \hat{\vc{y}}
      - \frac{1}{\tau_{\rm f}}\left(\vc{w}-\vc{u}\right)\, .
  \label{eq:dusteqmot}
\end{eqnarray}
Transport terms on the left hand side of \Eqs{eq:dustcontinuity}{eq:dusteqmot}
include advection by the peculiar velocities, $\vc{w}$, and by the Kepler
shear, $\vc{V}_0$. The right hand side of the equation of motion (Eq.\
[\ref{eq:dusteqmot}]) contains Coriolis forces (as modified by Kepler shear)
and drag acceleration relative to the gas component with velocity $\vc{u}$. We
apply a linear drag force with constant friction time $\tau_{\rm f}$, valid for
relatively small particles in the Epstein or Stokes regimes \citep{ahn76,
Weidenschilling1977}. Epstein's Law, $\tau_{\rm f}^{\rm (Ep)} = \rho_\bullet
R/(\rho_{\rm g} c_{\rm s})$ holds for particles of size $R \lesssim
\lambda_{\rm g}$, where $\lambda_{\rm g} \approx (r/{\rm AU})^{2.75}\, {\rm
cm}$ is the mean free path of the gas molecules, $c_{\rm s}$ is the gas sound
speed, and $\rho_\bullet$ is the internal density of rock/ice. Stokes' Law,
$\tau_{\rm f}^{\rm (St)} = \tau_{\rm f}^{\rm (Ep)} R/\lambda_{\rm g}$ applies
in the relatively narrow range $\lambda_{\rm g} \lesssim R \lesssim
\lambda_{\rm g} v_{\rm K}/c_{\rm s}$, where $v_{\rm K} \equiv \varOmega r$ is
the local Keplerian speed. Yet larger particles, $R \gtrsim \lambda_{\rm g}
v_{\rm K}/c_{\rm s}$, trigger turbulent wakes with non-linear drag
accelerations, which can not be modeled with a constant friction
time.\footnote{The onset of turbulent wakes would be stalled to larger
particles if the relative velocity $|\vc{u} - \vc{w}| < \eta v_{\rm K}$. In
practice, however, particles this large are weakly coupled and experience the
full pressure-supported headwind.} Note that Stokes' Law is independent of gas
density (since $\lambda_{\rm g} \propto 1/\rho_{\rm g}$). The dependence of
Epstein's law on gas density fluctuations is neglected in our calculations as
it is a small correction for low Mach number flow.

The solid component does not feel a pressure gradient, neither from the gas,
because the mass per solid particle is so high, nor from interparticle
collisions, because the number density is so low. Drag effects dominate
collisional effects, since the collision time, $t_{\rm coll} = \rho_\bullet
R/(\rho_{\rm p} c_{\rm p})$, is long with $t_{\rm coll}/\ts \approx (\rho_{\rm
g}/\rho_{\rm p})( c_{\rm s}/c_{\rm p}) \gg 1$, even when the particle density
is large, since the rms speed of particles, $c_{\rm p}$, is much smaller than
the gas sound speed.\footnote{If the particles were large enough for non-linear
turbulent drag, then collisional effects could only be safely neglected for
$\rho_{\rm g} > \rho_{\rm p}$ and/or if drift motions dominate particle random
motions. With $\ts^{\rm (turb)} \approx \rho_\bullet R/(\rho |\vc{u} -
\vc{w}|)$, then $t_{\rm coll}/\ts^{\rm (turb)} \approx (\rho_{\rm g}/\rho_{\rm
p})( |\vc{u} - \vc{w}|/c_{\rm p})$.}

For numerical work, we also use a Lagrangian description of particle motion,
see \S\ref{s:solidsaspar}.

\subsubsection{Gas Evolution}

The equations of continuity and motion for the gas read
\begin{eqnarray}
  \frac{\dpa \rho_{\rm g}}{\dpa t} + \vc{u}\cdot \nab \rho_{\rm g}
      - \frac{3}{2} \varOmega x \frac{\dpa \rho_{\rm g}}{\dpa y}
     & = & - \rho_{\rm g} \nab \cdot \vc{u} \, ,
  \label{eq:continuity}\\
  \frac{\partial \vc{u}}{\dpa t} + (\vc{u}\cdot\nab)\vc{u}
      - \frac{3}{2} \varOmega x \frac{\dpa \vc{u}}{\dpa y} 
      &=& 2 \varOmega u_y \hat{\vc{x}} \nonumber \\
      & & \hspace{-1.0cm}- \frac{1}{2} \varOmega u_x \hat{\vc{y}}
      - c_{\rm s}^2 \nab \ln \rho_{\rm g} \nonumber \\
      & & \hspace{-1.0cm} +  2 \eta  \varOmega^2 r \hat{\vc{x}}
      - \frac{\ep}{\tau_{\rm f}}\left(\vc{u}-\vc{w}\right)
  \, . \label{eq:gaseqmot}
\end{eqnarray}
Equation (\ref{eq:continuity}) reduces to $\nab \cdot \vc{u} = 0$ for an
incompressible gas, as was considered in YG. The momentum equation
\begin{figure*}[!t]
   \includegraphics[width=1.0\linewidth]{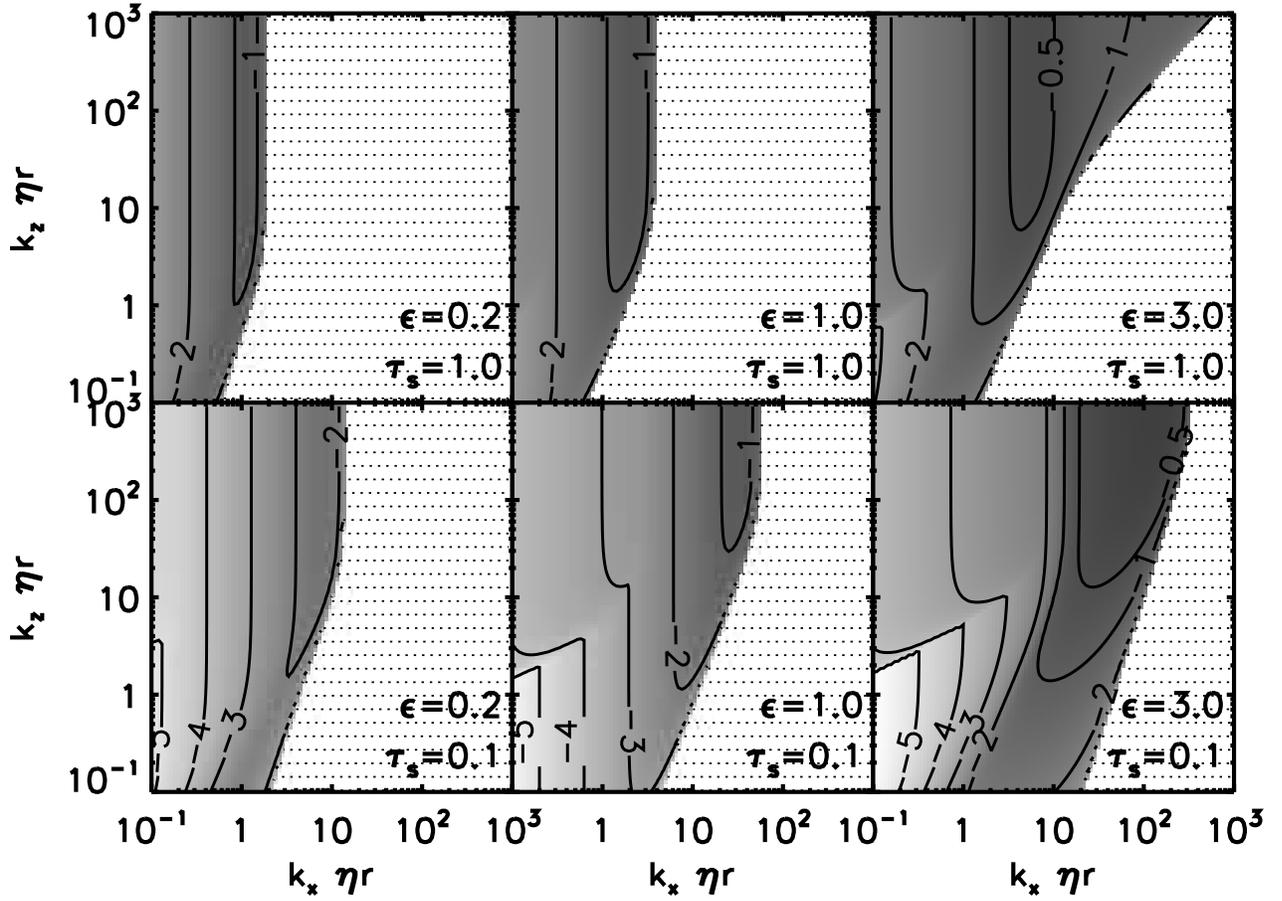}
   \caption{Linear growth rate $s$ of the streaming instability vs.\ radial and
     vertical wavenumbers for a friction time of $\tau_{\rm s}=1.0$ (upper row)
     and $\tau_{\rm s}=0.1$ (lower row). Three values of the solids-to-gas
     density ratio, $\epsilon=0.2,1.0,3.0$, are considered along the columns.
     Contours label $\log_{10}(s/\varOmega)$, darker shading corresponds to
     faster growth rates, while the dotted regions contain only damped modes.}
  \label{f:growthrates}
\end{figure*}
(\ref{eq:gaseqmot}) contains advection and Coriolis forces as
\Eq{eq:dusteqmot}. The main distinction between the two components is that gas
is effected by pressure gradients. We include both local pressure gradients
from isothermal gas density fluctuations and a constant acceleration by a
global radial pressure gradient,\footnote{While seemingly paradoxical, it is
consistent with the local approximation to include global pressure gradients to
linear order while treating the background gas density as constant.} $\dpa
P/\dpa r$, expressed using the dimensionless measure of sub-Keplerian rotation
\begin{equation}
  \eta \equiv - {\dpa P / \dpa r \over 2 \rho_{\rm g} \varOmega^2 r} \sim \frac{c_{\rm s}^2}{v_{\rm K}^2}\, .
\end{equation}
The feedback of the linear drag force scales with the density ratio of
particles to gas,
\begin{equation} 
  \epsilon \equiv \rho_{\rm p}/\rho_{\rm g} \, ,
\end{equation} 
which ensures that total momentum is conserved.

\subsection{Equilibrium State}\label{s:equil}

Equilibrium solutions to the mutually coupled \Eqs{eq:dusteqmot}{eq:gaseqmot}
were obtained by \citet[][ hereafter referred to as NSH]{Nakagawa+etal1986} for
local and linear dynamics. The in-plane deviations from Keplerian rotation are
\begin{eqnarray}
  u_x &=& \,\,\, \frac{2 \epsilon \tau_{\rm s}}
      {(1+\epsilon)^2+\tau_{\rm s}^2} \eta v_{\rm K}
      \, , \label{eq:NSHux} \\
  u_y &=& -\left[1+ \frac{\ep \tau_{\rm s}^2}
      {(1+\epsilon)^2+\tau_{\rm s}^2} \right]\frac{\eta v_{\rm K}}{1+\ep}
      \, , \label{eq:NSHuy} \\
  w_x &=& -\frac{2 \tau_{\rm s}}
      {(1+\epsilon)^2+\tau_{\rm s}^2} \eta v_{\rm K}
      \, , \label{eq:NSHwx} \\
  w_y &=& -\left[1- \frac{ \tau_{\rm s}^2}
      {(1+\epsilon)^2+\tau_{\rm s}^2} \right]\frac{\eta v_{\rm K}}{1+\ep}
      \, . \label{eq:NSHwy}
\end{eqnarray}
The dimensionless stopping time, $\tau_{\rm s} \equiv
\varOmega \tau_{\rm f}$, is a convenient measure of coupling strength, since
marginal coupling, $\tau_{\rm s} =1$, famously maximizes the radial drift speed
of an isolated particle. Velocities scale with the sub-Keplerian velocity,
$\eta v_{\rm K}$, where $v_{\rm K} \equiv \varOmega r$.\footnote{The radial
dependence of equilibrium quantities, such as $\eta v_{\rm K}$ and $\tau_{\rm
s}$, is ignored in the local approximation since the effect of the Keplerian
shear profile is already included (to linear order) in the equations of
motion.} The azimuthal velocities are factored into the center-of-mass motion,
\begin{equation} 
  V_y^{\rm (com)} \equiv
      \frac{\rho_{\rm g} u_y + \rho_{\rm p} w_y}{\rho_{\rm p}+\rho_{\rm g}}
      = -\frac{\eta v_{\rm K}}{1+\epsilon}\, ,
\end{equation} 
and order $\tau_{\rm s}^2$ drift motions (see YG for details).

Vertical gradients in the solids-to-gas ratio $\epsilon$ give gradients in
$V_y^{\rm (com)} \approx u_y \approx w_y$ (for $\tau_{\rm s} \ll 1$) that
trigger the settling-induced Kelvin-Helmholz instabilities discussed in the
introduction. As in YG, we also neglect vertical gravity in the present work in
order to allow for a laminar equilibrium state. With vertical gravity, any
initial condition must be time-dependent (due to vertical settling) and/or
turbulent (to halt the settling). Furthermore, in stratified disks, drift
speeds (and even directions) vary with height above the midplane, since
$\tau_{\rm s}$ rises with decreasing gas density and since the radial gas
pressure gradient can reverse away from the mid-plane \citep{tak02}. This is
particularly relevant for small grains that remain above the midplane for many
orbital times. The severity of the unstratified approximation is justified by
the insights gained from an initially simple, well-defined problem that rapidly
turns complex.

\subsection{Streaming Instability}\label{s:SIanalytic}

The streaming motion of solid particles through gas presents a source of free
energy that is driven by pressure gradients and mediated by drag and Coriolis
forces. YG showed, by linearly perturbing \Eqs{eq:dusteqmot}{eq:gaseqmot} about
the equilibrium state given by \Eqss{eq:NSHux}{eq:NSHwy}, that this streaming
robustly triggers instability in protoplanetary disks. The instability provides
a novel mechanism to generate growing particle density perturbations in a
moderately dense mid-plane layer of macroscopic particles, while smaller
particles ($\tau_{\rm s} \ll 1$) with poor drag feedback ($\epsilon \ll 1$)
will give rise to only very low, sub-dynamical growth rates.

The YG analysis and the linear test simulations in this paper are ``2.5-D'',
i.e. all three components of velocity fluctuations are considered,\footnote{And
all three components are necessary for axisymmetric instability (YG).} but
perturbations are axisymmetric and characterized by the radial and vertical
wavenumbers, $k_x$ and $k_z$. The growth rates for several choices of
$\tau_{\rm s}$ and $\epsilon$ (which henceforth indicates the \emph{average}
value of $\rho_{\rm p}/\rho_{\rm g}$ in the background state, unless otherwise
noted) are shown in \Fig{f:growthrates} as a function of the dimensionless
wavenumbers $K_x \equiv k_x \eta r$ and $K_z = k_z \eta r$. 

\begin{figure}
  \includegraphics[width=8.7cm]{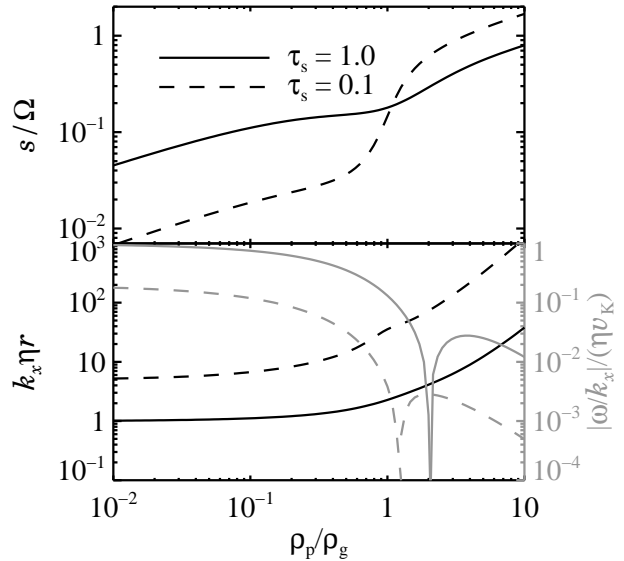}
  \caption{Peak growth rate, $s$, of the streaming instability and fastest
    growing radial wavenumber, $k_x$, versus the solids-to-gas density ratio
    $\epsilon = \rho_{\rm p}/\rho_{\rm g}$ for a friction time of $\tau_{\rm
    s}=1.0$ (solid line) and $\tau_{\rm s}=0.1$ (dashed line). Growth becomes
    faster and occurs at smaller scales for increasing $\ep$, with a
    particularly sharp increase in $s$ across $\ep = 1$ for tightly coupled
    particles with $\tau_{\rm s}=0.1$. Gray curves in lower plot (associated
    with gray axis on right) show the radial phase speed of waves. The sharp
    dips near $\ep \approx 1$--$2$ indicate a sign change for the wave speed:
    inward when gas dominates and outward when particles dominate.}
  \label{f:KzInf}
\end{figure}
Since particles only affect gas dynamics via drag feedback, growth rates
increase for larger $\epsilon$, while the relevant length scales shrink, most
likely because the response time-scale of the gas speeds up as $\ts/\epsilon$.
\Fig{f:KzInf} shows these trends, along with the particularly sharp increase of
$s$ across $\epsilon = 1$ for tightly coupled particles with $\tau_{\rm
s}=0.1$. The crucial physical distinction for marginal coupling (for which the
same sharp increase is not present) may be that for $\tau_{\rm s} \approx 1$,
azimuthal drift (of order $\tau_{\rm s}^2$) is no longer negligible compared to
radial drift (of order $\tau_{\rm s}$). For a more technical difference, note
the gray curves in \Fig{f:KzInf}, which show that the phase speed of waves
changes sign near $\epsilon \approx 1$. YG noted that the phase speed tends to
track the component with the fastest radial drift -- solids for $\ep < 1$ and
gas for $\ep > 1$. Curiously at $\tau_{\rm s} = 1$ the transition is delayed to
$\ep \simeq 2$. As $\tau_{\rm s}$ decreases the switch in phase speeds gets
closer to $\ep = 1$, coinciding with the rise in growth rates across $\ep = 1$
becoming steeper and of larger amplitude (see also Fig. 3 of YG for the
$\tau_{\rm s} = 0.01$ case). 
\begin{deluxetable*}{lllllllll}
  \tablecaption{Test Mode Eigensystems}
  \tablewidth{0pt}
  \tablehead{
    \colhead{} &
    \colhead{$\tilde{u}_x$} & \colhead{$\tilde{u}_y$} & \colhead{$\tilde{u}_z$}&
    \colhead{$\tilde{\rho}_{\rm g}$}&
    \colhead{$\tilde{w}_x$} & \colhead{$\tilde{w}_y$} & \colhead{$\tilde{w}_z$}&
    \colhead{$\omega$} }
  \startdata
    linA: $\tau_{\rm s}=0.1,\epsilon=3.0$ &
        $-0.1691398$ &
        $+0.1336704$ &
        $+0.1691389$ &
        $+0.0000224$ &
        $-0.1398623$ &
        $+0.1305628$ &
        $+0.1639549$ &
        $-0.3480127$ \\
    \quad($K_x=30,K_z=30$)  &
        $+0.0361553\ii$ &
        $+0.0591695\ii$ &
        $-0.0361555\ii$ &
        $+0.0000212\ii$ &
        $+0.0372951\ii$ &
        $+0.0640574\ii$ &
        $-0.0233277\ii$ &
        $+0.4190204\ii$ \\
    linB: $\tau_{\rm s}=0.1,\epsilon=0.2$ &
        $-0.0174121$ &
        $+0.2767976$ &
        $+0.0174130$ &
        $-0.0000067$ &
        $+0.0462916$ &
        $+0.2739304$ &
        $+0.0083263$ &
        $+0.4998786$ \\
    \quad($K_x=6,K_z=6$)     &
        $-0.2770347\ii$ &
        $-0.0187568\ii$ &
        $+0.2770423\ii$ &
        $-0.0000691\ii$ &
        $-0.2743072\ii$ &
        $+0.0039293\ii$ &
        $+0.2768866\ii$ &
        $+0.0154764\ii$ \\
  \enddata
  \tablecomments{Frequency $\omega$ is normalized to $\varOmega$, velocities
    are normalized to $\eta v_{\rm K}$, and densities to the average value for
    particles or gas respectively. All eigenvalue coefficients are relative to
    the particle density perturbation, which should be set to
    $\tilde{\rho}_{\rm p} \ll 1$ for the evolution of the mode to be linear.
    We used  $\tilde{\rho}_{\rm p} = 10^{-6}$ to normalize the eigenvector.
    The (tiny) effect of compressibility is included in the coefficients with
    $\eta v_{\rm K}/c_{\rm s} = 0.05$. The growth rate $s$ is the imaginary
    part of $\omega$.}
  \label{t:modes}
\end{deluxetable*}

The trend with $\tau_{\rm s}$ is complicated as well. In the gas-dominated
regime ($\epsilon < 1$) growth rates show the expected rise toward the
$\tau_{\rm s} \approx 1$ ``sweetspot": streaming motions are large yet
particles still respond effectively to the gas. The situation reverses when
particles dominate ($\epsilon > 1$), with growth rates that are actually faster
for tighter coupling, but at smaller length scales.

Returning for a moment to \Fig{f:growthrates}, it is also evident that growth
does not peak at a single pair of wavenumbers. The fastest growing $K_x$ can be
determined, with only damped modes for sufficiently large $K_x$, but growth
remains flat for large $K_z$ (indeed the curves of \Fig{f:KzInf} are calculated
in the limit $K_z/K_x \gg 1$). A physical explanation for the difference
between large or small $K_z/K_x$ follows. The (near) incompressibility of the
gas imposes a ratio $|u_z/u_x| \simeq |K_x/K_z|$. With $K_z \gg K_x$, velocity
vectors are nearly parallel to the $x-y$ plane with negligible vertical
velocities (just enough to maintain gas incompressibility). Since the balance
of Coriolis forces is maintained in thin vertical sheets, instability persists
to large $K_z$. On the other hand, large $K_x/K_z$ shrinks $u_x$ and destroys
the necessary balance of Coriolis forces.

The linear growth regime is surprisingly complex, considering the simplicity of
the physical system. Toy models to explain the mechanism have unfortunately
fallen short of capturing the essence of the instability. For instance, one
might suspect that, since streaming instabilities involve particle density
enhancements, they arise because radial drift slows in overdense regions [see
\Eq{eq:NSHwx}] leading to local traffic jams.\footnote{Not to be confused with
the global pile-ups that arise for Epstein drag either without \citep{ys02} or
with \citep{yc04} drag feedback.} This effect, while relevant, does not explain
linear growth of infinitesimal perturbations. To see this, consider the
axisymmetric evolution of particle density that follows from the equilibrium
drift speed [\Eq{eq:NSHwx}] and continuity [\Eq{eq:dustcontinuity}], which we
express for simplicity in terms of a \emph{variable} (only for now) $\epsilon =
\rho_{\rm p}(x,t)/\rho_{\rm g,0} = \epsilon_0 + \epsilon'(x,t)$ as
\begin{equation} \label{eq:toydrift}
\frac{\dpa \epsilon'}{\dpa t} = - \frac{\dpa (\epsilon w_x)}{\dpa x} = 2\eta v_{\rm K} \tau_{\rm s}\frac{\dpa }{\dpa x}\left[\frac{\epsilon}{(1+\epsilon)^2 + \tau_{\rm s}^2}\right]\, .
\end{equation} 
Linearizing about $\epsilon' \ll \epsilon_0$ clearly gives stable wave
propagation at the drift speed. Non-linear perturbations in \Eq{eq:toydrift}
will steepen a particle density wave, with no amplitude growth \citep[readily
shown by the method of characteristics, see][]{shuv2}. Even if the traffic jam
concept fails to explain the linear growth of the streaming instability, it may
be used to explain the non-linear clumping seen in JY (see also
\S\ref{s:Eclump} in this paper).

We find in JY that non-linear states also show remarkable diversity with
friction time and solids-to-gas ratio. We must, however, first ensure that the
numerical algorithms can capture and confirm the linear growth phase. 

\subsubsection{Eigenvectors and Vertical Standing Waves}

To test the growth rates of \Fig{f:growthrates} computationally, the
eigenvectors, i.e. relative amplitudes and phases of the density and velocity
perturbations, must be carefully seeded for a specific choice of parameters
$\tau_{\rm s}, \epsilon, K_x, K_z$. The perturbation in each dynamical variable
$f$ can be written in terms of its complex amplitude $\tilde{f}$ (a component
of the full eigenvector) as $f(x,z)=\Re\{\tilde{f}\exp[\ii(k_x x + k_z z -
\omega t)]\}$, where $\omega \equiv \omega_\Re + \ii s$ is the complex
eigenvalue containing the wave frequency $\omega_\Re$ and the growth rate $s$.
We choose to eliminate the superfluous vertical phase speed by superposing
pairs of modes with vertical wavenumbers $k_z$ and $-k_z$, respectively. Under
a vertical parity transformation the vertical velocity amplitudes are odd,
while all others are even. The superposition yields
\begin{eqnarray}
  f_{\rm e}(x,z) &=&\hspace{0.29cm}
      [\Re(\tilde{f}) \cos(k_x x - \omega_\Re t) - \nonumber \\
                 & & \hspace{-0.5cm} \Im(\tilde{f})
                     \sin(k_x x - \omega_\Re t)] \cos(k_z z)\exp(st)
  \label{eq:feven} \, , \\
  f_{\rm o}(x,z) &=&-
      [\Re(\tilde{f}) \sin(k_x x - \omega_\Re t) + \nonumber \\
                & & \hspace{-0.5cm} \Im(\tilde{f})
                    \cos(k_x x- \omega_\Re t)] \sin(k_z z) \exp(st) \, ,
  \label{eq:fodd}
\end{eqnarray}
for even (${\rm e}$) and odd (${\rm o}$) dynamical variables,
respectively, which are now clearly standing waves in $z$.

\Tab{t:modes} lists eigenvalues and eigenvectors for the cases we will test
numerically in \S\ref{s:linear}. The calculation is similar to that of YG
except gas compressibility was added so that a gas density perturbation can be
included in the numerical calculations. The effect of the gas compressibility
is otherwise negligible for $\eta v_{\rm K}/c_{\rm s} \sim c_{\rm s}/v_{\rm K}
\ll 1$ (the reason it was neglected in YG), affecting eigenvalues and
eigenvectors in the 5th digit for our choice of $\eta v_{\rm K}/c_{\rm s} =
0.05$. We also checked that the sound waves introduced by gas compression are
rapidly damped. Note that \Tab{t:modes} shows the gas density (and thus
pressure) perturbations are out of phase (by $\sim 90^\circ$ and $\sim
180^\circ$ for A and B, respectively) with the particle density perturbation.
Thus solids are not merely collecting in pressure maxima, as occurs in gas
density structures that are steady in time.

\section{Numerical Methods} \label{s:numerics}

As a numerical solver we use the Pencil Code.\footnote{The code is publicly
available at\\\url{http://www.nordita.dk/data/brandenb/pencil-code/}.} This is a
modular finite difference code that uses 6th order symmetric spatial
derivatives and a 3rd order Runge-Kutta time integration \citep[see][ for
details]{Brandenburg2003}. A module already exists for solving the equation of
motion of a dust fluid that interacts with the main gas fluid through drag
force \citep{jab04,JohansenKlahr2005}. The basic dynamical equations in the
(here unstratified) shearing sheet are \Eqs{eq:continuity}{eq:gaseqmot} for the
gas and \Eqs{eq:dustcontinuity}{eq:dusteqmot} for the solids. This equation set
is stabilized by adding small diffusive terms to the equation of motion and by
upwinding the advection term in the continuity equations \citep[for details,
see][]{JohansenKlahr2005,Dobler+etal2006}. Treating particles as a fluid
facilitates analytic calculations and is significantly cheaper for numerical
simulations, but is not always the desired approach.

\subsection{Solids as Particles}\label{s:solidsaspar}

Using Lagrangian particles provides a more realistic description of the
dynamics of the solids, and there are two main reasons to justify the
additional effort.\footnote{See \citet{gbl04} for a thorough analysis of the
validity of fluid descriptions of particle motion subject to gas drag} First,
particles at a given position need not have a single well-defined velocity as
the fluid approximation assumes, i.e.\ trajectories can cross. This concern is
particularly valid for marginal and looser coupling. Second, and more
seriously, the fluid treatment cannot capture large density gradients,
especially since the ``sound speed" of the pressureless fluid is zero.
Stabilization of steep density gradients would require a large artificial
viscosity that compromises the dynamics. Thus a Lagrangian treatment of the
solids is necessary for the non-linear simulations of JY which generate large
particles overdensities. Since the analysis of YG describes solids as a fluid,
we must demonstrate that the instability does not depend crucially on this
assumption.

When treating solids as numerical particles, or rather as {\it superparticles}
since each numerical particle effectively represents a huge number of
individual solids, each particle $i$ has a position $\vc{x}^{(i)}$ and a
velocity $\vc{v}^{(i)}$ relative to the Keplerian shear. Particle motions are
governed by Hill's equations \citep{wt88}
\begin{eqnarray}
  \frac{\de \vc{v}^{(i)}}{\de t} &=& 
      2 \varOmega v_y^{(i)} \hat{\vc{x}}
      - \frac{1}{2} \varOmega v_x^{(i)} \hat{\vc{y}} \nonumber \\
      & & \hspace{2.5cm}
      - \frac{1}{\tau_{\rm f}} 
      \left[\vc{v}^{(i)}-\overline{\vc{u}(\vc{x}^{(i)})}\right] \, ,
      \label{eq:dustpeqmot}\\
  \frac{\de \vc{x}^{(i)}}{\de t} &=& \vc{v}^{(i)} -{3 \over 2} \varOmega
  x^{(i)} \hat{\vc{y}}\, ,
      \label{eq:dustpeqpos}
\end{eqnarray}
here including drag force and expressed in a form to appear as the Lagrangian
equivalent to \Eq{eq:dusteqmot}. For axisymmetric simulations in the
radial-vertical plane, the evolution of $v_y^{(i)}(t)$ is included but the
azimuthal component of \Eq{eq:dustpeqpos} is irrelevant, effectively replaced
by $\de y^{(i)}/\de t=0$ since that dimension that is not present. The
interpolation of gas velocities at the particle positions,
$\overline{\vc{u}(\vc{x}^{(i)})}$, is addressed in the next section.

\subsection{Drag Force Calculation}\label{s:dragcalc}

The computation of drag forces between Lagrangian particles and an Eulerian
grid requires some care to avoid spurious accelerations and to ensure momentum
conservation. Small errors in the gas velocity are dangerously amplified by the
subtraction of highly correlated particle velocities. Our drag force algorithm
involves three steps:
\begin{enumerate}
  \item Interpolating gas velocities at particle positions
  \item Calculating the drag force on particles
  \item Assigning the back-reaction force to the gas from particles in nearby
      cells
\end{enumerate}

For the first step, interpolation, we begin with gas velocities,
$\vc{u}^{(\vcs{j})}$, defined on a uniform grid where the index $\vc{j}$ labels
the cells centered on positions $\vc{x}^{(\vcs{j})}$. We interpolate to the
particle positions, $\vc{x}^{(i)}$, using a weight function, $W_{\rm I}$, as
\begin{equation}\label{InterpGasVel}
  \overline{\vc{u}(\vc{x}^{(i)})} = \sum_{\vcs{j}} W_{\rm I}(\vc{x}^{(i)} -
  \vc{x}^{(\vcs{j})})\vc{u}^{(\vcs{j})} \, .
\end{equation} 
The weight function is normalized as $\sum_{\vcs{j}} W_{\rm
I}(\vc{x}^{(i)}-\vc{x}^{(\vcs{j})}) = 1$ for any $\vc{x}^{(i)}$, and has
non-zero contributions only from the cells in the immediate vicinity of
$\vc{x}^{(\vcs{j})}$.

The second step, calculating the drag acceleration on particle $i$,
\begin{equation}\label{fpart}
  \vc{f}_{\rm p}^{(i)} = -{\vc{v}^{(i)} - \overline{\vc{u}(\vc{x}^{(i)})} \over \ts} \, ,
\end{equation}
is trivial once the relevant quantities are defined, but this is the step that
amplifies interpolation errors in $\overline{\vc{u}(\vc{x}^{(i)})}$, because of
strong coupling to particle velocities, a problem that worsens for smaller
$\tau_{\rm f}$. We note that other choices of the drag law (e.g.\ non-linear in
the velocity or including gas density fluctuations in Epstein drag) would be
simple to implement by interpolating the relevant grid-based quantities as in
\Eq{InterpGasVel}. 

Finally, we calculate the back-reaction drag force, $\vc{f}_{\rm
g}^{(\vcs{j})}$, on the gas in cell $\vc{j}$. Assigning particle velocities to
a mesh risks violating momentum conservation. Instead we follow the suggestion
of Jim Stone (personal communication) and use Newton's third law to directly
assign the force on the particles back to the gas,
\begin{eqnarray} \label{eq:fgj}
  \vc{f}_{\rm g}^{(\vcs{j})} = -{m_{\rm p} \over \rho_{\rm g}^{(\vcs{j})} V_{\rm cell}} \sum_iW_A(\vc{x}^{(i)}-\vc{x}^{(\vcs{j})})\vc{f}_{\rm p}^{(i)} \label{newtonIII}\, ,
\end{eqnarray}
where $m_{\rm p}$ is the mass of a particle (if not uniform it would be inside
the sum), and $V_{\rm cell}$ is the volume of a grid cell. The assignment
function $W_{\rm A}$ obeys the same conditions as $W_{\rm I}$, so that only
particles in a given cell or its nearby neighbors contribute to the sum. Global
momentum conservation follows trivially from summation of \Eq{eq:fgj},
\begin{equation}
  V_\mathrm{cell} \sum_{\vcs{j}} \rho_\mathrm{g}^{(\vcs{j})} f_{\rm
  g}^{(\vcs{j})} + m_{\rm p}\sum_i f_p^{(i)} = 0\, ,
\end{equation}
with no reference to the drag law, the interpolation function, or any
properties of the assignment function except normalization. Thus unlike
particle-mesh calculations with interacting particles (e.g.\ by self-gravity),
we are flexible to choose $W_{\rm I}$ and $W_{\rm A}$ independently, without
violating momentum conservation. Nevertheless, choosing $W_{\rm A} = W_{\rm I}$
is safest since drag forces from gas to particles -- and vice-versa -- are
smoothed symmetrically.

We opted for second order interpolation and assignment methods, either
quadratic spline or quadratic polynomial, which use three grid cells in each
dimension, for a total of 9 (27) for 2-D (3-D) simulations, respectively. This
gave considerable improvement over lower order bilinear interpolation (but at a
computational cost -- the drag force calculations dominate the wall time in our
simulations with high order interpolation and assignment). The details and
errors associated with the interpolation schemes are described in Appendix
\ref{s:Werrors}. The quadratic spline assignment/interpolation method is often
referred to as the Triangular Shaped Cloud scheme \citep[TSC,
see][]{HockneyEastwood1981}.

\subsubsection{Boundary Conditions for the Drag Force}

Our implementation of periodic boundary conditions, and use of higher
\citep[than zeroth, as in][]{jhk06} order assignment schemes, causes particles
near grid edges to exert drag forces on mesh points across the boundaries. In
non-axisymmetric simulations (such as the 3-D simulations that we present in
JY) the radial direction is shear-periodic so that two connected points at the
inner and outer radial boundary are $\Delta y(t)={\rm mod}[(3/2) \varOmega L_x
t,L_y]$ apart in the azimuthal direction. Techniques for implementing radial
boundary conditions in the shearing box are well-known \citep{Hawley+etal1995}.
Fluid variables in zones on one radial boundary are copied to ghost zones
adjacent to the opposite boundary and shifted azimuthally. Then differences
across boundaries are performed, i.e.\ ``copy, shift, and difference." 

\begin{figure}
  \includegraphics[width=8.7cm]{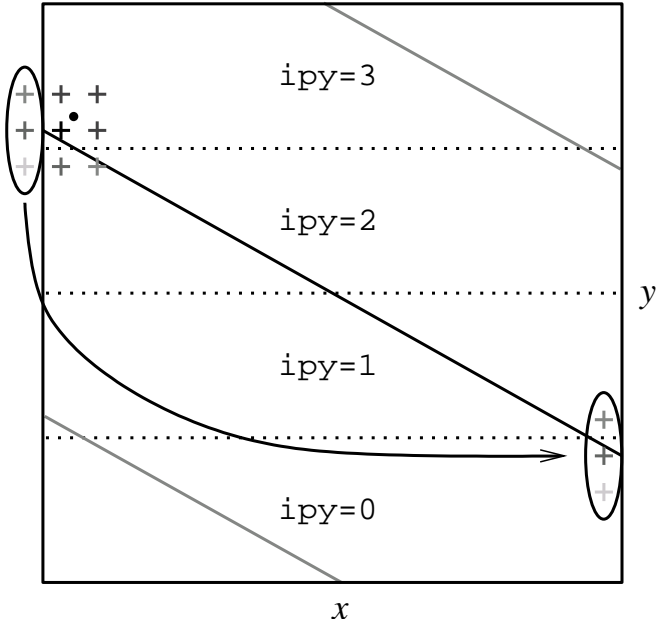}
  \caption{A sketch of the shear-periodic radial ($x$) boundary condition for
    the assignment of drag forces from a particle to the gas. The dot
    represents a particle near the boundary and crosses indicate the (centers
    of) gas cells that receive a drag acceleration with the second order TSC
    assignment scheme (grayscale of crosses indicates rough weight of drag
    force received by gas in each cell). We illustrate an example with 4
    processors in the $y$-direction (labeled \texttt{ipy}). The periodic
    direction is indicated by solid diagonal lines. The drag force assigned to
    ghost cells across the boundary (circled on left) is shifted in Fourier
    space and then added as an acceleration on the physical grid cells at the
    outer boundary. Note that in practice (a) the drag force from an
    individual particle influences more than three grid cells across the
    boundary, since displacements are not integer multiples of the grid spacing
    and (b) drag forces from all particles on a ghost zone are added before
    Fourier shifting.}
  \label{f:fold}
\end{figure}
The implementation of shear periodic boundary conditions for drag forces on the
gas is a subtly different ``assign, shift, and add" procedure, as sketched in
\Fig{f:fold}. First we assign the (appropriate fraction of) drag accelerations
from particles in boundary zones to gas in the ghost zones. Then we shift the
accelerations on the radial ghost zones in the $y$-direction, the inner by
$-\Delta y(t)$, and the outer by $+\Delta y(t)$. Finally these shifted
accelerations are added (or folded) to the first real zone on the opposite side
of the mesh. We interpolate (since the ghost zones do not slide by integer
numbers of grid cells) by applying the azimuthal shift in Fourier space.
Fourier interpolation has the advantage over high order polynomial
interpolation that the function and all its derivatives are continuous. A
numerical test of the radial boundary condition with shearing waves is
described in Appendix \ref{s:shearing}.

\section{Numerical Tests of Linear Growth}\label{s:linear}

We now present measurements of linear growth rates of the streaming instability
from numerical simulations. These results confirm the capabilities of our code
and verify the authenticity of this fundamental instability, not yet explicitly
established for a particle-based treatment of solids. Our efforts in
reproducing growth rates to a satisfactory accuracy were useful in developing
our numerical implementation of drag forces. We hope that others who simulate
coupled particle-gas disks will conduct similar dynamical tests of the simplest
(identified) aerodynamic drag instability.\footnote{AY will provide initial
conditions (eigenvectors) by e-mail request.}

We choose two different test problems: an eigenvector for $\tau_{\rm s}=0.1$,
$\epsilon=3.0$, $K_x=K_z=30$ (run linA), which grows rapidly with
$s/\varOmega=0.41902$, and an eigenvector for $\tau_{\rm s}=0.1$,
$\epsilon=0.2$, $K_x=K_z=6$ (run linB) that grows more slowly with
$s/\varOmega=0.01548 $ and hence is more numerically demanding. The total
initial velocities are the sum of the equilibrium drift solutions of
\Eqss{eq:NSHux}{eq:NSHwy}, and the vertically standing wave of
\Eqss{eq:feven}{eq:fodd} with eigenvectors from \Tab{t:modes}. The initial
amplitude of the particle density was set to $10^{-6}$ in all cases to ensure
linearity.

\subsection{Growth for Solids as a Fluid}

\begin{figure*}
  \includegraphics[width=\linewidth]{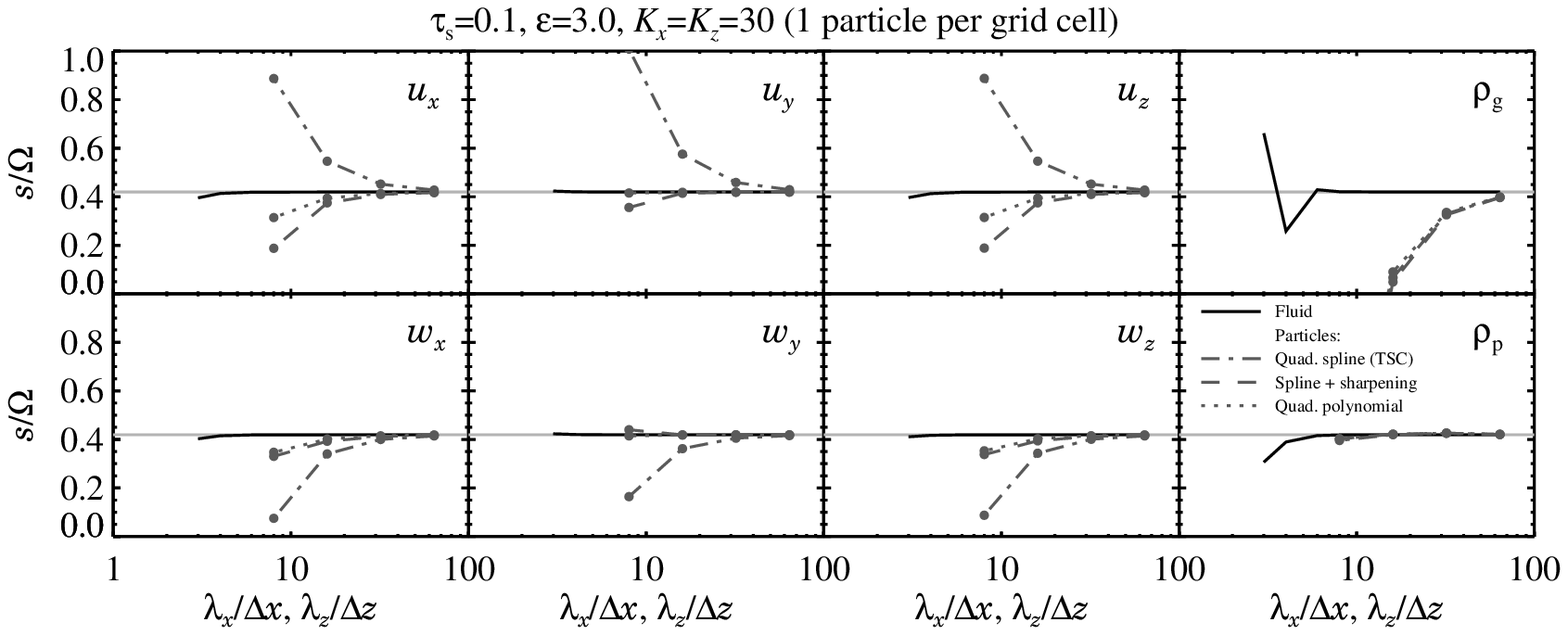}
  \includegraphics[width=\linewidth]{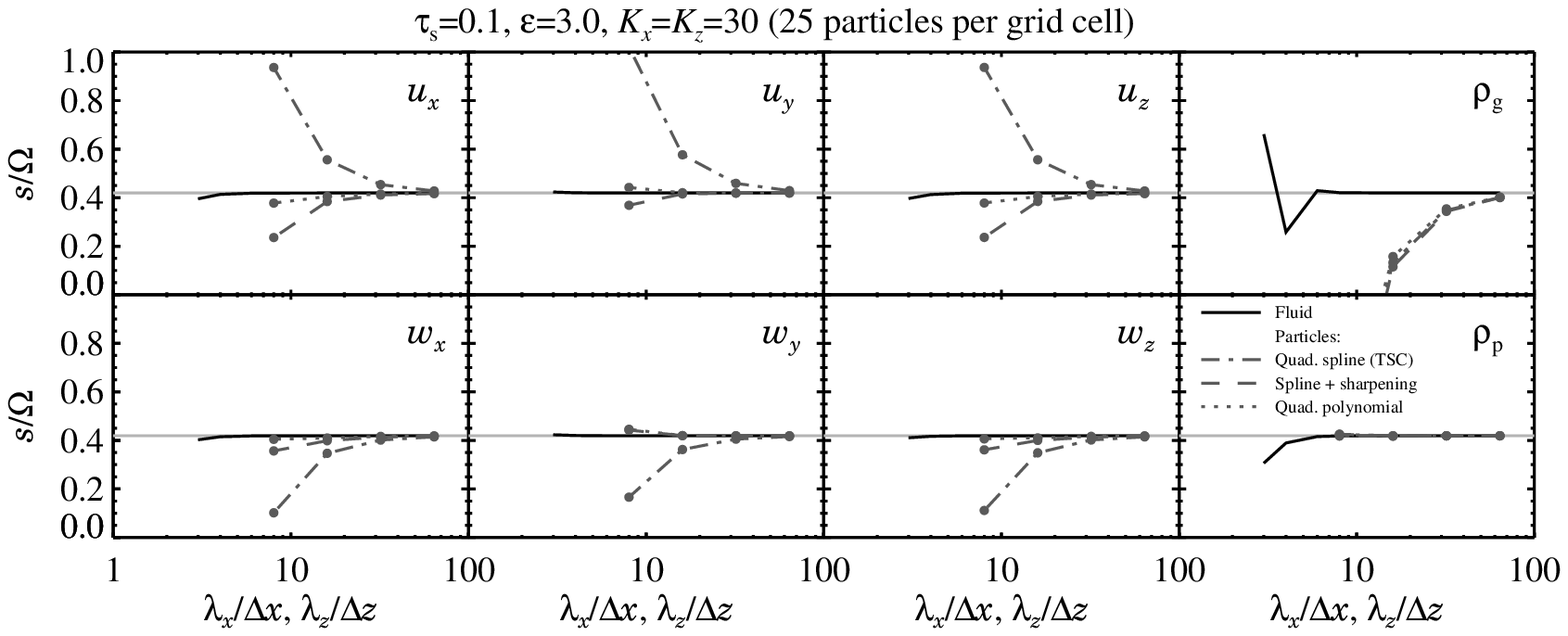}
  \caption{Measured growth rate of a seeded mode with $K_x=K_z=30$ and
    $\tau_{\rm s}=0.1$, $\epsilon=3.0$ as a function of the number of
    grid-points per wavelength, shown for 1 and 25 particles per grid cell in
    the top and bottom plots, respectively. The behavior of each dynamical
    variable is shown separately. The analytical growth rate,
    $s=0.41903\varOmega$, is indicated with a gray line. The fluid treatment
    (solid black line) gives excellent agreement with the analytical growth
    rate down to 4 grid points per wavelength, whereas 16 grid points is needed
    for the regular TSC scheme (dash-dotted line). Applying Fourier sharpening
    to the initial condition gives some improvement (dashed line). Replacing
    spline interpolation with polynomial interpolation (dotted line) gives
    better growth rates, but polynomial interpolation has the disadvantage of
    being discontinuous over cell interfaces. Increasing the number of
    particles per grid cell from 1 to 25 has minimal influence on the
    linear growth.}
  \label{f:growth_gridpoints_e3.0}
\end{figure*}
The measured growth rate when particles are treated as a fluid is shown with a
solid black line in \Figs{f:growth_gridpoints_e3.0}{f:growth_gridpoints_e0.2}
(the top and bottom plots are identical for the two-fluid case). The eight
panels show the growth rate of the velocity and density of the gas (top row)
and of the solids (bottom row) as a function of the number of grid points per
wavelength. We have varied the resolution between 3 and 64 grid points per
wavelength for the fluid treatment of solids and between 8 and 64 grid points
per wavelength for the particle treatment. The growth rates are obtained by
spatially Fourier transforming the 8 dynamical variables at 10 fixed times over
$\Delta t = 0.2 \varOmega^{-1}$ and measuring the amplitude growth of the
relevant Fourier mode. There is generally an excellent agreement between the
measured growth rates when the solids are treated as a fluid and the analytical
values down to 4 grid points per wavelength, except for the gas density which
shows some variation from the analytical value for crude resolutions. This
disagreement is not surprising since small errors in the cancellation of $\dpa
u_x/\dpa x$ and $\dpa u_z/\dpa z$ for the nearly incompressible gas give
spurious growth to the gas density according to the linearized continuity
equation $\dpa \ln \rho_{\rm g}'/\dpa t = - \nab \cdot \vc{u}'$. While the gas
density perturbations are too small to affect the drag force, they also cause
the pressure perturbations which are significant. Fortuitously, the errors in
the gas density (for crude resolutions) do not affect the other dynamical
variables. It may help that spurious sound waves damp rapidly (in a stopping
time).
\begin{figure*}
  \includegraphics[width=\linewidth]{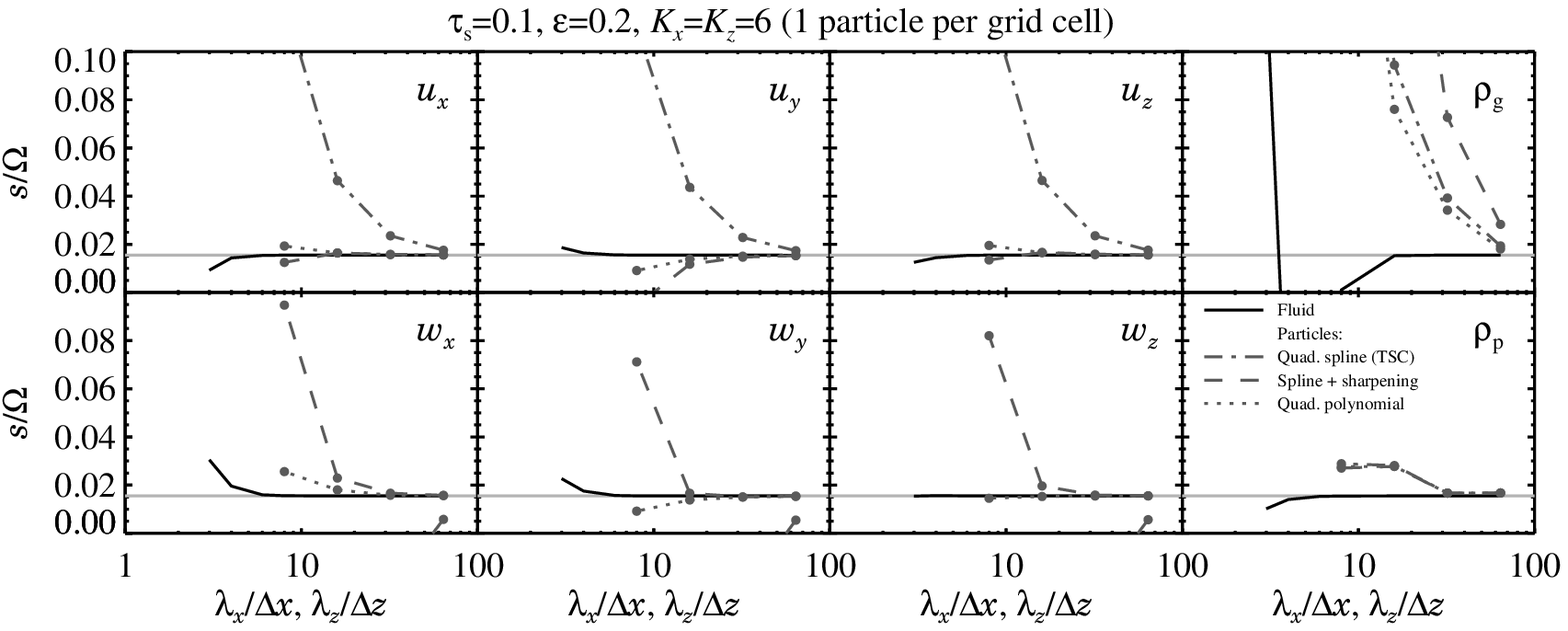}
  \includegraphics[width=\linewidth]{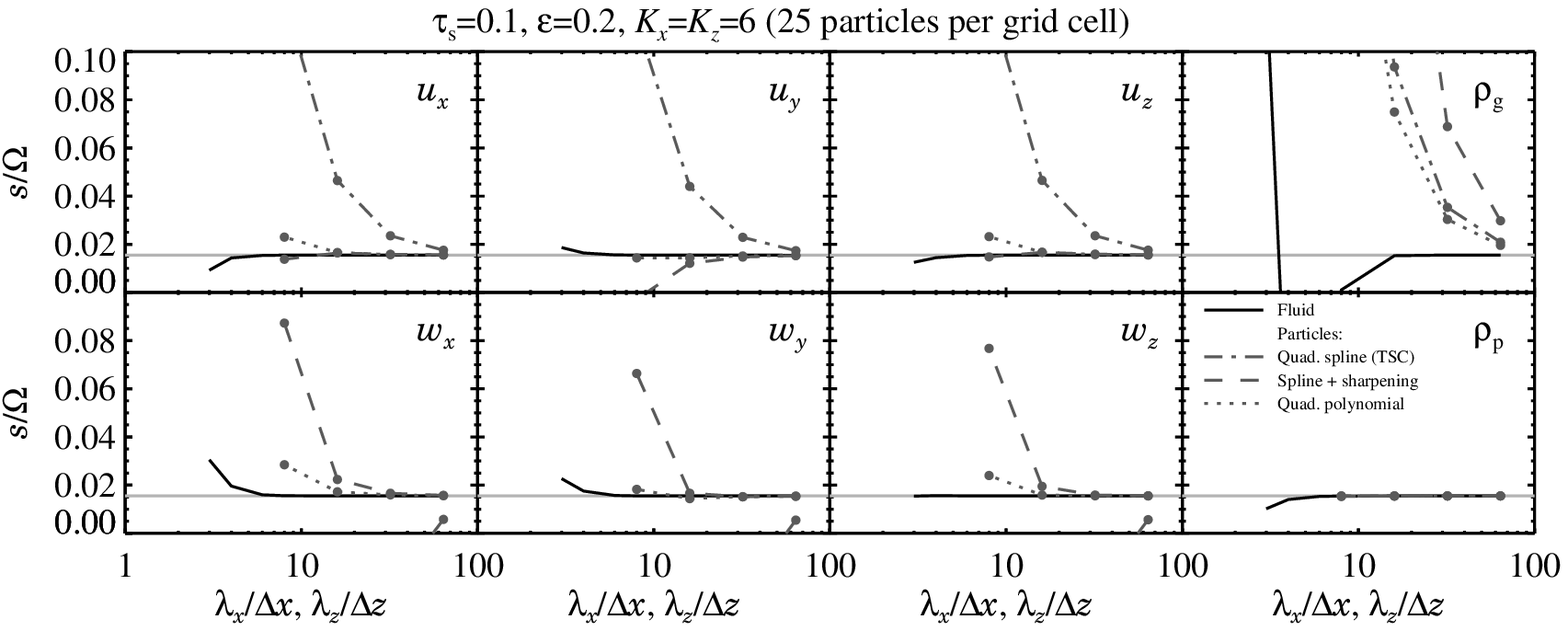}
  \caption{Similar to \Fig{f:growth_gridpoints_e3.0} but for a mode with
    $K_x=K_z=6$ and $\tau_{\rm s}=0.1$, $\epsilon=0.2$ that has an analytical
    growth rate of $s=0.01548\varOmega$. The agreement with the analytical
    growth shows a comparable resolution dependance to
    \Fig{f:growth_gridpoints_e3.0}, but here the increase to 25 particles per
    grid cell shows better agreement for $\rho_{\rm p}$.}
  \label{f:growth_gridpoints_e0.2}
\end{figure*}

\subsection{Growth for Solids as Particles}\label{s:linearpart}

Reproducing analytic growth rates using a particle representation of the solids
is significantly more difficult than in the two-fluid case. Poisson
fluctuations from undersampling and truncation errors in the drag force
calculation cause numerical discrepancies. Section \ref{s:dragcalc} and
Appendix \ref{s:Werrors} describe the algorithms for computing drag forces and
the errors associated with interpolation and assignment.

\subsubsection{Cold Start Initialization}

To avoid shot noise in seeding linear particle density perturbations we use a
``cold start'' algorithm (described in detail in Appendix \ref{s:coldstart})
for the initial particle positions. First we place all particles on a uniform
grid. Then we apply a small, spatially periodic shift to their positions. This
seeds the desired mode with minimal noise leaked to other wavelengths. We
experimented with different numbers of particles: 25 particles per grid cell to
match the non-linear runs of JY, and 1 particle per grid cell as a test.

\begin{figure}
  \includegraphics[width=8.7cm]{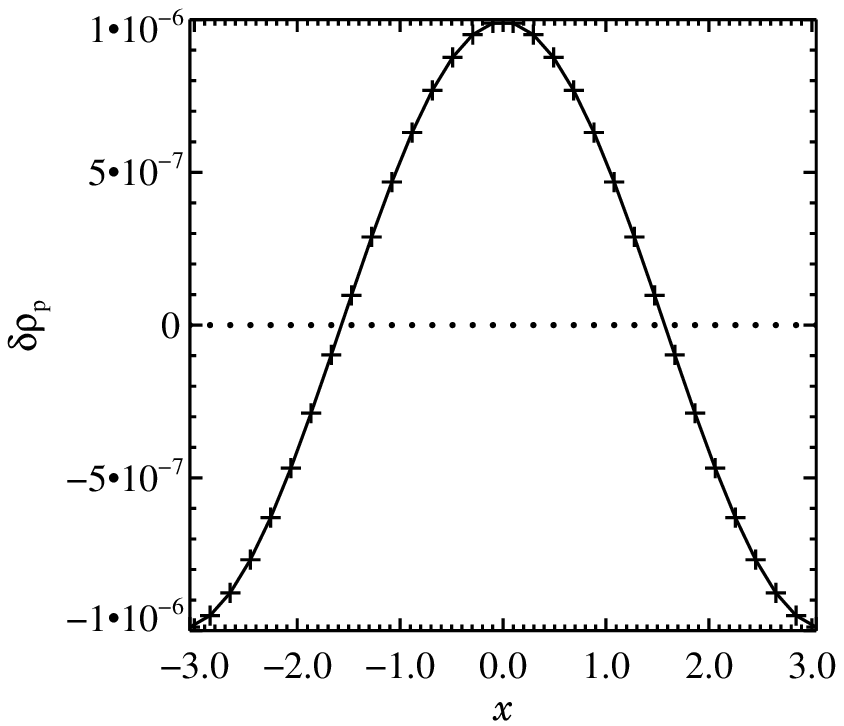}
  \caption{A sinusoidal particle density perturbation of amplitude $10^{-6}$ as
    generated by the shifting algorithm of Appendix \ref{s:coldstart} with 32
    particles -- only one per grid cell! The crosses (connected by the solid
    line) plot the TSC assignment of particle (over)density to the grid cells.
    Dots indicate the positions of the particles, but the shift is
    imperceptibly small.} 
  \label{f:amplitude}
\end{figure}
With the cold start to eliminate noise and the TSC assignment scheme to
smoothly distribute a particle's influence over the nearest three grid cells
per dimension, communicating initial density perturbations of infinitesimal
amplitude with only a few particles is trivial. \Fig{f:amplitude} demonstrates
the algorithm effectiveness with the near perfect replication of a 1-D particle
density perturbation of amplitude $10^{-6}$ with only 32 grid cells and one
particle per cell. This is nothing more (or less) than the miracle of
continuous numbers. The use of many particles per grid cell is still necessary
to get good statistics in non-linear simulations.

\subsubsection{Results}

The growth rates with solids as particles are shown (together with the
two-fluid results) in \Figs{f:growth_gridpoints_e3.0}{f:growth_gridpoints_e0.2}
as a function of spatial resolution. The top and bottom plots in each figure
are for 1 and 25 particles per grid cell, respectively. Particle number makes
little difference for the agreement with linear theory, although additional
particles give some improvement, notably for the growth rate of $\rho_{\rm p}$
in \Fig{f:growth_gridpoints_e0.2}.

While all runs use the TSC scheme to assign drag forces to the gas, three
different techniques were tested for the interpolation of gas velocities to
particle positions: (1) quadratic spline interpolation, (2) quadratic spline
interpolation with an initial Fourier sharpening of the gas velocity field, and
(3) quadratic polynomial interpolation. Errors in gas velocity interpolation
are the most dangerous since they are amplified in the force calculation by
subtracting a particle velocity that is highly correlated with the gas flow.

The first technique, quadratic spline interpolation, uses the same weight
function as TSC assignment and gives smooth interpolates with a reduced
fluctuation amplitude. The dash-dotted lines in
\Figs{f:growth_gridpoints_e3.0}{f:growth_gridpoints_e0.2} show that this
technique accurately reproduces the growth of $\rho_{\rm p}$. The results for
the other variables are poor for resolutions of less than 16 grid points per
wave length. This is a result of spurious drag forces generated because
interpolation reduces gas fluctuation amplitudes.\footnote{This is why, for
unsharpened spline interpolation, growth rates are too large for $\vc{u}$ (gas
is accelerated toward the unsmoothed amplitude by particles) and too small for
$\vc{w}$ (particles are decelerated by the lowered gas amplitudes).}

The second interpolation technique (shown with dashed lines in
\Figs{f:growth_gridpoints_e3.0}{f:growth_gridpoints_e0.2}) still uses quadratic
splines, but sharpens the initial gas velocities to correct the drag force.
The amplitude of the Fourier modes $\tilde{\vc{u}}$ are increased by the
precise amount, $[1-\Delta^2 (k_x^2+k_z^2)/8]^{-1}$, that interpolation reduces
them (see Appendix \ref{s:Werrors}). The sharpened TSC scheme gives much better
growth rates, but still not as good as the two-fluid results. In a non-linear
simulation with an evolving power spectrum, one could sharpen $\vc{u}$ with a
pair of Fourier transforms at each time-step, but this was deemed too
computationally costly. By getting improved results with only the initial
condition sharpened, we show that growth rate discrepancies with spline
interpolation are largely due to differences between numerical (discretized)
and analytic eigenvectors that should not compromise the non-linear simulations.

The third approach (shown with dotted lines in
\Figs{f:growth_gridpoints_e3.0}{f:growth_gridpoints_e0.2}) opts for precise
quadratic polynomial interpolation instead of smoother splines. The resulting
growth rates are comparable, or slightly better than, the sharpened splines.
Despite the simplicity and good results obtained with this technique, we did
not use it in the non-linear runs. Discontinuities in the interpolates at cell
boundaries would add noise by leaking power to the grid scale. Since the errors
of TSC are well-behaved (spatially smooth across a grid cell, declining with
increasing resolution, and leaving particle density growth unaffected even at
low resolution), we used spline interpolation in the non-linear runs. We also
prefer the symmetry of using the same weight functions for interpolation
(quadratic spline) and assignment (TSC).

Overall, numerical growth rates with solids treated as particles agree well
with linear theory down to 16 grid points per wavelength, although the particle
density grows at the correct rate even at 8 grid points per wavelength.
Anomalies, particularly in the gas density, suggest that sound waves are being
triggered due to interpolation errors, but these spurious motions damp and do
not impede the expected growth of particle density perturbations.

\section{Energy and Angular Momentum Balance}\label{s:EL}

This section provides brief overviews of energy and angular momentum in coupled
particle-gas disks in order to provide a point of reference to more familiar
dynamical systems, and because it will help us interpret the non-linear results
of JY. We denote $\mathcal{L} \equiv \rho_{\rm g}u_y + \rho_{\rm p}w_y$ as the
total angular momentum density of solids and gas, ignoring the radius factor
that is constant in the local approximation. The azimuthal components of
\Eqs{eq:dusteqmot}{eq:gaseqmot} give
\begin{equation} \label{eq:Ldot}
  \left({\dpa \over \dpa t}
      - {3 \over 2} \varOmega x {\dpa \over \dpa y}\right) \mathcal{L}
      + \nab \cdot \vc{\mathcal{F}}_{\mathcal{L}}
      = - {\varOmega \over 2}\mathcal{F}_{\rho, x} - {\dpa P \over \dpa y} \, .
\end{equation}
The terms on the left hand side relate local changes in $\mathcal{L}$ to the
transport of $\mathcal {L}$ by the Keplerian flow and to the angular momentum
flux $\vc{\mathcal{F}}_{\mathcal {L}} \equiv \rho_{\rm g} u_y \vc{u} +
\rho_{\rm p} w_y \vc{w}$. We do not call this flux a Reynolds stress because
the velocities $\vc{u}$ and $\vc{w}$ have not been decomposed into fluctuations
about their mean. The NSH equilibrium of \Eqss{eq:NSHux}{eq:NSHwy} transports
angular momentum radially inwards,
\begin{eqnarray} 
  \mathcal{F}_{\mathcal{L},x} &\equiv&
      \rho_{\rm g} u_x u_y + \rho_{\rm p} w_x w_y \\
      &=&-2 \tau_{\rm s}^3 \rho_{\rm p}
      \left[ \frac{\eta v_{\rm K}}{(1+\epsilon)^2+\tau_{\rm s}^2} \right]^2
      \label{eq:FLx}\, ,
\end{eqnarray} 
a consequence of the slower rotation of the outgoing gas
relative to the faster rotation of the incoming particles. This differs from
the usual outward transport of angular momentum in accretion disks
\citep{lp74}, because the driving agent is not orbital shear, but the radial
pressure gradient.

The terms on the right hand side of \Eq{eq:Ldot} represent sources or sinks of
angular momentum: the radial mass flux, $\mathcal{F}_{\rho, x} \equiv \rho_{\rm
g} u_x + \rho_{\rm p}w_x$, and azimuthal pressure gradients, where $P$ is
promoted to denote the total gas pressure (background and perturbations) in
this section. Equation (\ref{eq:Ldot}) proves that axisymmetric equilibrium
solutions cannot transport mass radially in the local model, a condition obeyed
by \Eqs{eq:NSHux}{eq:NSHwx}. Note that \Eq{eq:Ldot} does not explicitly include
drag forces, which transfer momentum between gas and solids, but (of course) do
not dissipate $\mathcal{L}$.

The evolution of kinetic energy density $\mathcal{E} \equiv (\rho_{\rm g}
|\vc{u}|^2 + \rho_{\rm p} |\vc{w}|^2)/ 2$ is found by summing the dot products
of $\rho_{\rm g}\vc{u}$ with \Eq{eq:gaseqmot} and $\rho_{\rm p}\vc{w}$ with
\Eq{eq:dusteqmot} to give
\begin{equation}
  \left({\dpa \over \dpa t}
      -{3 \over 2} x\varOmega{\dpa \over \dpa y}\right) \mathcal{E} + \nab \cdot
      \vc{\mathcal{F}}_{\mathcal{E}}
      = \dot{\mathcal{E}}_\mathrm{drag} -\vc{u} \cdot \nab P
      + {3 \over 2}\varOmega \mathcal{F}_{\mathcal{L},x} \, ,
\end{equation}
where the energy flux, $\vc{\mathcal{F}}_{\mathcal{E}} \equiv \rho_{\rm g}
|\vc{u}|^2 \vc {u} + \rho_{\rm p} |\vc{w}|^2 \vc{w}$, transports energy
radially inward (outward) when gas (particles) dominate the mass,
respectively.\footnote{Actually this result only holds in the center of mass
reference frame, i.e.\ with $V_y^{\rm (com)}$ subtracted.} The sources and
sinks on the right hand side include the energy lost to drag dissipation,
\begin{eqnarray}
  \dot{\mathcal{E}}_\mathrm{drag} &\equiv&
      -\rho_{\rm p} |\vc{w}-\vc{u}|^2/ \ts \\
      &=&-{4(1+\ep)^2 \tau_{\rm s}
      +  \tau_{\rm s}^3 \over [(1+\ep)^2 + \tau_{\rm s}^2]^2}(\eta v_{\rm K})^2\rho_{\rm p}\varOmega\, , \label{eq:dEdragEq}
\end{eqnarray}
where the second equality applies to the NSH equilibrium. A
simple estimate of the effective temperature produced when the dissipated
kinetic energy is released as thermal heat gives
\begin{equation} 
  T_{\rm drag} < \left[\varSigma_{\rm p} \varOmega (\eta v_{\rm K})^2/\sigma_{\rm
  SB}\right]^{1/4} \sim 30 (r/\mathrm{AU})^{-3/4}\, \mathrm{K}
\end{equation} 
as an upper limit for the case of marginal coupling and $\ep \ll 1$, where
$\varSigma_{\rm p} \simeq \rho_{\rm p} H_{\rm p}$ is the surface density of the
solid component and $H_{\rm p}$ is the scale height of the sublayer of solids.
The above temperature limit is significantly colder than even passively
irradiated disks \citep{cg97}, a comforting fact for SED modelers.

The $\dot{\mathcal{E}}_{\rm work} \equiv - \vc{u}\cdot \nab P$ term represents
energy gained from the work done by the total pressure forces. The equilibrium
value of
\begin{equation} 
  \dot{\mathcal{E}}_{\rm work} = - u_x (\dpa P/\dpa r) = {4 \tau_{\rm s} \over
(1+\ep)^2 + \tau_{\rm s}^2}(\eta v_{\rm K})^2\rho_{\rm p}\varOmega \label{eq:dEwork}
\end{equation} 
shows that $|\dot{\mathcal{E}}_{\rm work}| >
|\dot{\mathcal{E}}_\mathrm{drag}|$, i.e.\ more energy is put into the system by
pressure work than removed by drag. The final term,
\begin{equation} 
  \dot{\mathcal{E}}_{\mathcal{L}} \equiv {(3/ 2)}\varOmega \mathcal{F}_{\mathcal{L},x} = -{3 \tau_{\rm s}^3 \over
\left[(1+\ep)^2 + \tau_{\rm s}^2\right]^2}(\eta v_{\rm K})^2\rho_{\rm
p}\varOmega \, , \label{eq:dEL}
\end{equation}
is well known in studies of viscous or collisional disks as the heat generated
by the outward transport of angular momentum \citep{ss85,lc06}. However, in our
case angular momentum transport is reversed according to \Eq{eq:FLx} and
provides a sink of kinetic energy. The phenomenon of ``backwards" angular
momentum transport, and the dynamical cooling it provides, has been famously
offered as an explanation for the sharp edges of planetary rings \citep{bgt82}. 

Equations (\ref{eq:dEdragEq}), (\ref{eq:dEwork}), and (\ref{eq:dEL}) verify
that the heating and cooling terms sum up to zero in the equilibrium state:
$\dot{\mathcal{E}}_{\rm drag}+\dot {\mathcal{E}}_{\rm
work}+\dot{\mathcal{E}}_{\mathcal{L}} = 0$. The work done by pressure forces
balances dissipation by drag forces and losses from the backwards transport of
angular momentum. 

\subsection{Clumping and Dissipation} \label{s:Eclump}

In this subsection we will show that particle clumping reduces energy
dissipation by drag forces, at least in a laminar state. Particles effectively
``draft" off each other like birds flying in formation or bicycle riders in a
peloton. This drafting does not rely on overlapping turbulent wakes, but
instead depends on slowing relative gas motions by the collective inertia of
particles. It is tempting to argue that the lowered dissipation rate explains
the tendency of particles to clump. As usual, the story is more complicated,
but the evolution of $\dot{\mathcal{E}}_{\rm drag}$ turns out to be a useful
diagnostic for the non-linear simulations of JY. 

First we demonstrate that dissipation is reduced by clumping. Consider the
equilibrium drag dissipation of \Eq{eq:dEdragEq}, for simplicity in the tight
coupling limit ($\tau_{\rm s} \ll 1$), which we now express per unit surface
area instead of volume as
\begin{equation}
  \Lambda_{\rm diss} \equiv \dot{\mathcal{E}}_\mathrm{diss} H_{\rm p} \approx - {4 \tau_{\rm s}  \over (1+\ep)^2}(\eta
v_{\rm K})^2\varSigma_{\rm p} \varOmega \, .
\end{equation}
Now imagine concentrating the particles into a volume smaller by a factor $n >
1$ via vertical setting or clumping. Compared to the uniform solids-to-gas
ratio $\ep$ the new value is $n \ep$ in clumps and 0 in voids. The new
height-averaged dissipation rate is 
\begin{equation}
  \Lambda_\mathrm{diss}^* = -{4 \tau_{\rm s} \over (1+n \ep)^2}(\eta
v_{\rm K})^2\varSigma_{\rm p}\varOmega\, .
\end{equation}
The fractional change in dissipation (for $\tau_{\rm s} \ll 1$),
\begin{equation}
f_\Lambda \equiv {\Lambda_\mathrm{diss}^* \over
\Lambda_\mathrm{diss}}= \left({ 1+ \ep \over 1+n \ep}\right)^2 < 1\, ,
\end{equation}
shows that clumping decreases the net dissipation of well-coupled particles and
that the effect becomes stronger with increasing $\epsilon$.

Unfortunately there is no reason to expect in general that the dissipation rate
decreases, especially since the system is not closed, but driven by pressure
gradients. Examples of driven systems in which mechanical dissipation increases
with the spontaneous transition from laminar to turbulent flow include drag on
a rigid body (e.g.\ an airplane wing) and Rayleigh convection with fixed
temperature on the endplates (Jeremy Goodman, personal communication). Indeed
the non-linear simulations of JY find that $|\dot{\mathcal{E}}_{\rm drag}|$
could increase or decrease in the non-linear state. Obviously drag dissipation
is affected not just by clumping (as in the toy laminar calculation here) but
by the turbulent velocities that tend to increase dissipation. Nevertheless JY
demonstrate that runs with the largest (and longest lived) overdensities show a
decrease in $|\dot{\mathcal{E}}_{\rm drag}|$, lending credence to the
hypothesis that drafting can augment particles' ability to clump. 

\section{Discussion}\label{s:disc}

This paper begins our numerical exploration of the streaming instability, which
uses aerodynamic particle-gas coupling to tap the radial pressure gradient in
protoplanetary disks. Growing oscillations arise in an idealized model for
protoplanetary disks that assumes a local, unstratified, and
non-self-gravitating shearing box with gas and uniformly-sized, non-colliding
solids. Studying a relatively simple system isolates the surprisingly rich
consequences of mutual drag coupling in disks. Also, the well-defined growth
rates of seeded eigenvectors make the streaming instability an ideal test of
numerical implementations of particle-gas dynamics, as suggested in YG. We
encourage those who study manifestations of particle-gas dynamics in disks to
consider the linear streaming instability as a test problem if the feedback of
solids on gas dynamics is relevant. 

This work is largely successful in reproducing the analytic growth rates of
YG. The two-fluid simulations, which treat solids as a pressureless fluid,
give excellent results with minimal computational effort. Particle-fluid
simulations also converge to the analytic results, but higher spatial
resolution is required. Treating the solids as particles has several
advantages -- it is more realistic, it can validate the often-used fluid
approximation for solids, and it allows the development of non-linear density
enhancements without spurious shocks. Refinements of the particle-fluid
algorithm used in \citet{jhk06} are described, notably the use of higher order
interpolation and assignment schemes to minimize errors in the drag force
computation. These errors become more drastic as the stopping time decreases
and errors grow relative to the diminishing difference between gas and particle
velocities. Smaller stopping times also give shorter length scales, thereby
imposing stricter Courant criteria. These restrictions actually dominate the
obvious concern that tighter coupling stiffens the equations of motion.
Detailed modeling of the smallest particles in protoplanetary disks, especially
in the inner regions with high gas densities, will require further algorithm
development and increased computational power. In the meantime, studies of
moderate coupling can establish the relevant physical phenomena and provide a
baseline for extrapolation to more extreme parameters.

Having developed a particle-mesh scheme that can be trusted to simulate coupled
particle-gas dynamics with feedback, we proceed to explore the non-linear
evolution of streaming instabilities in a companion paper (Johansen \& Youdin
2007) with particular attention to the growth and saturation of particle
overdensities.

\acknowledgments

The authors would like to thank Jeremy Goodman and Jim Stone for valuable
advice and encouragement throughout this project. An anonymous referee is
thanked for a number of suggestions that helped improve the original
manuscript. A.Y. acknowledges support NASA grant NAG5-11664 issued through the
Office of Space science. A.J. is grateful to the Annette Kade Graduate Student
Fellowship Program for his stay at the American Museum of Natural History.

{\it Facilities:} \facility{RZG (PIA)}

\appendix

\section{Weight Functions and Interpolation/Assignment Errors}\label{s:Werrors}

The choice of weight functions for interpolation, $W_{\rm I}$, and assignment,
$W_{\rm A}$, involves trading computational cost against performance. We will
consider only 1-D weight functions which are combined multiplicatively to form
multidimensional weights, $W(\vc{x} - \vc{x}_{\vcs{j}}) = W(x -x_\ell)W(y -
y_m)W(z - z_n)$, for the cell centered at $\vc{x}_{\vcs{j}} = x_\ell
\hat{\vc{x}}+y_m\hat{\vc{y}}+z_m\hat{\vc{z}}$. This assumes a rectilinear
domain of influence, which is simpler (if less physical) than
circular/spherical clouds.

The interpolation function in the non-linear simulations uses quadratic splines
(QS),
\begin{equation}\label{TSCweight}
  W_{\rm I}^{\rm (QS)}(\delta x_\ell) = \left\{
  \begin{array}{ll}
    {3 \over 4} - {\delta x_\ell^2 \over \Delta^2} & \mbox{ if $|\delta x_\ell| < \Delta/2$}\\
    {1 \over 2}\left( {3 \over 2} - {|\delta x_\ell| \over \Delta}\right)^2 & \mbox{ if $\Delta/2< |\delta x_\ell| < 3 \Delta/2$}\\
    0 & \mbox{ if $|\delta x_\ell| > 3 \Delta/2$}
  \end{array}\right.
  \, ,
\end{equation}
where $\delta x_\ell \equiv x - x_\ell$ measures the distance from a cell
center and $W_{\rm I}$ extends over three cells of width $\Delta $. The
interpolation errors are calculated by considering a periodic function (of
arbitrary phase) sampled at the grid points. The QS interpolated values at
arbitrary $x$ in cell $\ell$ are
\begin{eqnarray} 
  \overline{\cos(k x + \phi)}_{\rm I}^{\rm (QS)} &\simeq& \cos(k x +
  \phi)\left[1-{(\Delta k)^2 \over 8} \right] +
  \sin(k x + \phi)\left[{(\Delta k)^3 \over 24} \delta x_\ell(1-4\delta x_\ell^2)
  \right]  +\mathcal{O}(\Delta k)^4 \, .
\end{eqnarray}
The amplitude of a periodic signal is reduced by $1- (\Delta k)^2/8$. For an
arbitrary distribution, this smoothing can be simply corrected for in Fourier
space. This sharpening is included, but only for the initial amplitudes of
$\vc{u}$, in some linear tests (\S\ref{s:linearpart}), but was too costly for
the non-linear runs in JY. There is also a noise, i.e.\ an error that depends
on position relative to cell center $\delta x_\ell$, of amplitude $(k
\Delta)^3/(72 \sqrt{3})$.

We also considered quadratic polynomial interpolation (QP), which performs a
best fit through the three nearest grid points, resulting in a weight function
\begin{equation}
  W_{\rm I}^{\rm (QP)}(\delta x_\ell) = \left\{
  \begin{array}{ll}
    1 - {\delta x_\ell^2 \over \Delta^2}, & \mbox{ if $|\delta x_\ell| < \Delta/2$}\\
    {1 \over 2}\left( {1} - {|\delta x_\ell| \over \Delta}\right)\left( {2} - {|\delta x_\ell | \over \Delta}\right),& \mbox{ if $\Delta/2< |\delta x_\ell| < 3 \Delta/2$}\\
    0, & \mbox{ if $|\delta x_\ell | > 3 \Delta/2$}
  \end{array}
  \right. \, .
\end{equation}
The QP interpolated values of a periodic signal read
\begin{equation} 
  \overline{\cos(k x + \phi)}_{\rm I}^{\rm (QP)} \simeq \cos(k x + \phi)
      + \sin(k x + \phi)\left[{(\Delta k)^3 \over 6} \delta x_\ell(1-\delta x_\ell^2)\right] + \mathcal{O}(\Delta k)^4\, .
\end{equation} 
The amplitude is preserved to second order, an improvement over quadratic
spline interpolation. The noise, however, has an amplitude of $(\Delta
k)^3/16$, a factor of 15 larger than with quadratic spline. Also troubling is
the discontinuity of $(\Delta k)^3/8$ at the cell boundaries.

The assignment function used in all the non-linear simulations of JY is the
Triangular Shaped Cloud (TSC) scheme, as opposed to the lower order NGP
\citep[Nearest Grid Point, which was used in][]{jhk06} or CIC (Cloud In Cell)
schemes \citep{HockneyEastwood1981}. The TSC assignment weight function is
identical to the quadratic spline interpolation weight function, $W_{\rm
A}^{\rm (TSC)} \equiv W_{\rm I}^{\rm (QS)}$. Assignment errors depend partly on
sampling, i.e.\ the number of particles that make a non-zero contribution to a
sum like \Eq{eq:fgj}. A higher order method like TSC samples more particles at
a given grid point and gives a smoother distribution than lower order methods.
The fractional amplitude reduction of a mode perfectly sampled by TSC is
identical to the result for quadratic spline interpolation, $1- (\Delta
k)^2/8$. This is the same order, but larger than for CIC [$1- (\Delta k)^2/12$]
or NGP [$1- (\Delta k)^2/24$] assignment, although especially the NGP scheme
would require an enormous particle number to achieve good sampling of linear
perturbations. In principle Fourier sharpening could be applied in the force
assignment step, but it is less important to consider, since the errors are not
magnified by a subsequent subtraction.

\section{Numerical Test of Particle Assignment over Shearing Boundaries} \label{s:shearing}

\begin{figure}
  \begin{center}
    \includegraphics[width=8.7cm]{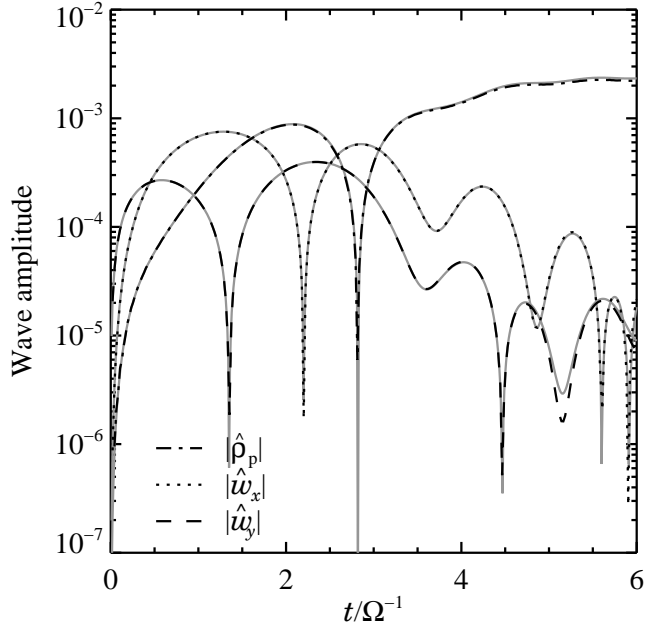}
  \end{center}
  \caption{The temporal evolution of a leading shear wave of gas and solid
    particles. The plot shows a comparison between the semi-analytic solution
    to the linearized equation system (gray lines) and the solution obtained
    with the full solver of the Pencil Code for the amplitude of the particle
    density $\hat{\rho}_{\rm p}$ (black dash-dotted line) and of the particle
    velocity components $\hat{w}_x$ (black dotted line) and $\hat{w}_y$ (black
    dashed line). There is excellent agreement between the numerical and the
    analytical solutions up until $t\simeq5\varOmega^{-1}$ where damping of the
    amplitude of the tightly wound wave by the Triangular Shaped Cloud scheme
    becomes significant.\vspace{0.4cm}}
  \label{f:testmode}
\end{figure}

This Appendix studies the behavior of a linear, non-axisymmetric wave of gas
and particles in order to test drag force assignment across the shear-periodic
radial boundary. Solutions from the full simulation of the Pencil Code are
compared to the following semi-analytic problem. We use local, linearized
equations of continuity and motion to describe the evolution of gas and
particle density $\rho_{\rm g}'(x,y,t)$, $\rho_{\rm p}'(x,y,t)$ relative to a
constant density background state $\rho_{{\rm g},0}$, $\rho_{{\rm p},0}$, and
gas and particle velocities $\vc{u}'(x,y,t)$, $\vc{w}'(x,y,t)$ relative to the
Keplerian shear flow $\vc{V}_0=-(3/2)\varOmega x \hat{\vc{y}}$.
We assume a shearing wave solution, $q'(x,y,t) = \hat{q}(t) \exp[\ii(k_x(t) x +
k_y y)]$, for each perturbation variable, with \citep[see e.g.][]{glb65II,
Brandenburg+etal2004}
\begin{equation} 
 k_x(t)=k_x(0)+(3/2) \varOmega t k_y\, .
\end{equation}
The wave amplitudes then evolve as coupled ordinary differential equations in
time\footnote{Note that for this test problem no global pressure gradient, and
thus no drift motions, are included.},
\begin{eqnarray}
\de \hat{\rho}_{\rm g}/\de t &=&
      -\rho_{{\rm g},0} [\ii k_x(t) \hat{w}_x + \ii k_y \hat{w}_y]
  \label{eq:lin_rhoghat} \, , \\
  \de \hat{u}_x/\de t &=&
       2 \varOmega \hat{u}_y
      -\ep_0
      (\hat{u}_x-\hat{w}_x)/\tau_{\rm f}
      - \ii k_x(t) c_{\rm s}^2\hat{\rho}_{\rm g}/\rho_{{\rm g},0}
  \label{eq:lin_uxhat} \, , \\
\de \hat{u}_y/\de t &=&
      - \varOmega \hat{u}_x/2
      -\ep_0
      (\hat{u}_y-\hat{w}_y)/\tau_{\rm f}
      - \ii k_y c_{\rm s}^2\hat{\rho}_{\rm g}/\rho_{{\rm g},0}
  \label{eq:lin_uyhat} \, , \\
\de \hat{\rho}_{\rm p}/\de t &=&
      -\rho_{{\rm p},0} [\ii k_x(t) \hat{w}_x + \ii k_y \hat{w}_y]
  \label{eq:lin_rhophat} \, , \\
\de \hat{w}_x/\de t &=&
       2 \varOmega \hat{w}_y
      - (\hat{w}_x-\hat{u}_x)/\tau_{\rm f}
  \label{eq:lin_wxhat} \, , \\
\de \hat{w}_y/\de t &=&
      - \varOmega \hat{w}_x/2
      - (\hat{w}_y-\hat{u}_y)/\tau_{\rm f}
  \label{eq:lin_wyhat} \, .
\end{eqnarray}
where $\ep_0 \equiv \rho_{{\rm p},0}/\rho_{{\rm g},0}$. We
solve this system of ordinary differential equations numerically for
$\varOmega=\rho_{{\rm g},0}=\rho_{{\rm p},0}=\tau_{\rm f}=c_{\rm s}=k_y=1$
using a third-order Runge-Kutta time integration method to follow the temporal
evolution of a non-axisymmetric wave with the initial condition $k_x=-1,
\hat{u}_x=\hat{\rho}_{\rm g}=\hat{w}_x=\hat{w}_y=\hat{\rho}_{\rm p}=0,
\hat{u}_y=10^{-3}$. The semi-analytic solution is then compared to the
evolution obtained with the full solver of the Pencil Code using $64^2$ grid
points with 1 particle per grid point to cover a box of size $L_x=L_z=2 \pi$.
\Fig{f:testmode} shows the evolution of the absolute value of the particle
amplitudes $\hat{\rho}_{\rm p}$ (dash-dotted line), $\hat{w}_x$ (dotted line)
and $\hat{w}_y$ (dashed line) in comparison with the analytical solution (gray
lines). There is an excellent agreement for $t\lesssim5.0$. At later times, the
wave becomes so tightly wound that damping of the wave amplitude by the TSC
scheme becomes significant. Most importantly this non-axisymmetric test problem
never shows any spurious features near the radial boundary (or anywhere else),
validating our implementation of drag force assignment over the boundaries.

\section{Cold Start: Algorithm for Seeding Density Perturbations}
\label{s:coldstart}

Seeding low amplitude (we use $\delta_{\rm p} = 10^{-6}$) density perturbations
with particles is non-trivial. The desired density distribution cannot be
seeded by random numbers for a reasonable number of particles, $N_{\rm p}$.
The white Poisson noise has a constant Fourier amplitude of $\sim
1/\sqrt{N_{\rm p}}$ at all scales, i.e., we would need a total number of
particles $N_{\rm p} \gg 10^{12}$ to resolve $\delta_{\rm p} = 10^{-6}$!

Instead we borrow a tactic from cosmological simulations \citep[e.g.][]{trac06}
to concentrate power in a desired mode. We first assign particles to a uniform
grid with positions, $\vc{x}_i$, labeled by a particle index $i =
1,2,...,N_{\rm p}$. This grid is defined relative to the gas grid with an
integer number of particles in each gas cell. We introduce linear perturbations
to the density by applying periodic shifts to the particle positions. To
approximate a density distribution
\begin{equation}
  \rho_{\rm p}(\vc{x}) = \langle\rho_{\rm p}\rangle[1+A \cos(\vc{k}_0\cdot\vc{x})]
\end{equation}
with $A \ll 1$, the desired shift from the uniform grid is
\begin{equation}
  \dv_i = -{\vc{k}_0 \over k_0^2}{A }\sin(\vc{k}_0\cdot\vc{x}_i)\, .
\end{equation}
The resulting density distribution,
\begin{equation}\label{eq:shiftdist}
  \rho_{\rm p}(\vc{x}) = \sum_i \delta (\vc{x} - \vc{x}_i - \dv_i)\, ,
\end{equation}
has a Fourier transform
\begin{eqnarray}
  \tilde{\rho}_{\rm p}(\vc{k}) 
  &=&V_{\rm cell}^{-1}\sum_i \exp\left\{\ii \vc{k}\cdot\left[\vc{x}_i +\dv_i \right]\right\}\\
  & \approx& {N_{\rm p} \over V_{\rm cell}}\left[\delta_{\vcs{k},0} + A(\delta_{\vcs{k},\vcs{k}_0}+\delta_{\vcs{k},-\vcs{k}_0})/2\right] + \tilde{\rho}_{\rm p}^{(2)}(\vc{k})+ \mathcal{O}(A^3)\, ,
\end{eqnarray}
where $V_{\rm cell}$ is the volume of a grid cell. The final
step shows that we reproduce the desired plane wave to lowest order by
performing an expansion about $\dv_i \ll \vc{x}_i$ and using summation
relations for periodic functions (we ignore the sub-gridscale aliases of
$\vc{k}_0$). The standing wave solutions described in \Sec{s:SIanalytic} are
produced by summing two plane-wave displacements of $\pm k_z$. The quadratic
error term
\begin{equation}
  \tilde{\rho}_{\rm p}^{(2)}(\vc{k}) = {N_{\rm p} \over V}{A^2 \over 2}(\delta_{\vcs{k},2 \vcs{k}_0}+\delta_{\vcs{k},-2\vcs{k}_0})
\end{equation}
is small for $A \ll 1$, but is eliminated by a further displacement
\begin{equation}
  \dv_i^{(2)} = {\vc{k}_0 \over 2k_0^2}{A^2 }\sin(2 \vc{k}_0\cdot\vc{x}_i)\, .
\end{equation}
Thus \Eq{eq:shiftdist} has the desired Fourier properties even with only one
particle per grid cell. However, the binned density distribution is actually
what is relevant for influencing gas dynamics. When the TSC assignment scheme
(described in Appendix B) is applied to the ``cold start" positions, we cleanly
get the desired density distribution assigned on the mesh, even for arbitrarily
small shift amplitudes.

\end{document}